\newif\iflongversion
\longversiontrue

\iflongversion
\documentclass[letterpaper,onecolumn]{IEEEtran}

\else
\documentclass[conference,letterpaper]{IEEEtran}

\fi

\usepackage[utf8]{inputenc} 
\usepackage[T1]{fontenc}
\usepackage{url}              % provides \url{...}
\usepackage{cite}             % improves presentation of citations

\usepackage[cmex10]{amsmath}  % Use the [cmex10] option to ensure complicance
\usepackage{mleftright}       % fix to wrong spacing of \left-,
\mleftright                   % \middle- \right-commands 

\usepackage{graphicx}         % provides \includegraphics{...} to
\usepackage{booktabs}         % fixes poor spacing in tables and
\usepackage{amssymb}
\usepackage{esint}

\makeatletter
\def\useAmsThmOrIEEE{0}
\ifnum\useAmsThmOrIEEE=1
\usepackage{amsthm}
\theoremstyle{plain}
\newtheorem{thm}{\protect\theoremname}
\theoremstyle{definition}
\newtheorem{defn}[thm]{\protect\definitionname}
\theoremstyle{plain}
\newtheorem{prop}[thm]{\protect\propositionname}
\theoremstyle{plain}
\newtheorem{lem}[thm]{\protect\lemmaname}
\theoremstyle{plain}
\newtheorem{cor}[thm]{\protect\corollaryname}
\theoremstyle{definition}
\newtheorem{example}[thm]{\protect\examplename}
\theoremstyle{definition}
\newtheorem{rem}[thm]{\protect\remarkname}
\else
\newtheorem{thm}{\protect\theoremname}
\newtheorem{prop}[thm]{\protect\propositionname}
\newtheorem{lem}[thm]{\protect\lemmaname}

\fi

\IEEEoverridecommandlockouts

\usepackage{mathrsfs}
\usepackage{algorithm,algpseudocode}

\usepackage{colortbl}
\definecolor{lightgray}{rgb}{0.9,0.9,0.9}
\definecolor{lightred}{rgb}{1,0.8,0.8}
\definecolor{lightgreen}{rgb}{0.6,1,0.6}
\definecolor{lightyellow}{rgb}{1,1,0.5}
\definecolor{lightgrey}{rgb}{0.8,0.8,0.8}

\allowdisplaybreaks[1]

\makeatother

\usepackage{babel}
\providecommand{\corollaryname}{Corollary}
\providecommand{\definitionname}{Definition}
\providecommand{\lemmaname}{Lemma}
\providecommand{\propositionname}{Proposition}
\providecommand{\theoremname}{Theorem}
\providecommand{\examplename}{Example}
\providecommand{\remarkname}{Remark}

\begin{document}
\title{Variable-Length Secret Key Agreement via Random Stopping Time}
\author{%
  \IEEEauthorblockN{Junda Zhou and Cheuk Ting Li\\}
  \IEEEauthorblockA{Department of Information Engineering, The Chinese University of Hong Kong, Hong Kong, China\\
                    Email: davidjdzhou@link.cuhk.edu.hk, ctli@ie.cuhk.edu.hk}
\thanks{
This work was partially supported by two grants from the Research Grants Council of the Hong Kong Special Administrative Region, China [Project No.s: CUHK 24205621 (ECS), CUHK 14209823 (GRF)].
}
}
\maketitle
\begin{abstract}
We consider a key agreement setting where two parties observe correlated random sources, and want to agree on a secret key via public discussions. In order to allow the key length to adapt to the realizations of the random sources, we allow the key to be of variable length, subject to a novel variable-length version of the uniformity constraint based on random stopping time. We propose simple, computationally efficient key agreement schemes under the new constraint. The proposed scheme can be considered as the key agreement analogue of variable-length source coding via Huffman coding, and the Knuth-Yao random number generator.
\end{abstract}
% \begin{IEEEkeywords}
% Secret key agreement, variable-length coding, random stopping time.
% \end{IEEEkeywords}

\medskip{}

\allowdisplaybreaks

\section{Introduction}

In the secret key agreement setting, Alice and Bob observe $X$ and $Y$ respectively, where $X, Y$ are sampled from a correlated source, and they wish to agree on a key as long as possible. This setting has been studied in the asymptotic setting where $X, Y$ are i.i.d. sequences with length approaching infinity. If no public discussion is allowed, the secret key rate was proven in \cite{gacsCommonInformationFar1973} to be much lower than the mutual information $I(X; Y)$. A series of work \cite{maurerSecretKeyAgreement1993, ahlswedeCommonRandomnessInformation1993, maurerUnconditionallySecureKey1999, maurerInformationTheoreticKeyAgreement2000} was devoted to studying the setting allowing public discussion, as well as correlated side information $Z$ available to the eavesdropper, where we also require the key to be kept secret from the eavesdropper. Asymptotic results on the secret-key rate under the strong secrecy requirement were proved in \cite{maurerInformationTheoreticKeyAgreement2000}. 

The key agreement setting has also been investigated in the one-shot and finite-blocklength setting, where each of $X, Y$ is a single symbol. One-shot and second-order bounds on the key length were studied in \cite{hayashiSecretKeyAgreement2016}. A one-shot converse bound was derived in \cite{tyagi2015converses}. While one-shot and asymptotic key agreement were traditionally studied in the fixed-length setting where the key length is fixed, a universal asymptotic key agreement result has been given in \cite{tyagi2017universal} where the key length can adapt to the distribution of the source sequence.

A one-shot variable-length key agreement scheme has been studied in~\cite{li2021one}, where the parties can agree on a key length $L$ that can adapt to the observed source symbols, but the key is required to be approximately uniformly distributed over $\{0,1\}^\ell$ (the set of bit sequences of length $\ell$) conditioned on $L=\ell$.
Compared to a fixed-length key, a variable-length key provides an improvement on the key length. It was proved in~\cite{li2021one} that the optimal expected key length is close to the mutual information $I(X; Y)$ within a logarithmic gap. 
While the conditional uniformity requirement ensures the quality of the key conditioned on any length, it is a stringent requirement that requires complicated schemes (involving spectrum slicing), and a considerable penalty on the key length. For example, even for the case where Alice and Bob observe a common random source symbol $X$, the scheme in~\cite{li2021one} can only generate a key of length $\approx H(X)- \log H(X) - 7$, with a logarithmic penalty from $H(X)$.

In this paper, we propose a new criterion for variable-length keys based on random stopping time, and study key agreement schemes under the new criterion. Compared to the conditional uniformity requirement~\cite{li2021one}, the new criterion is less stringent, and allows much simpler schemes achieving a longer average length. For example, for the case where Alice and Bob observe a common random source symbol $X$, the new scheme can generate a key of average length $\ge H(X)- 2$, with only a constant gap from $H(X)$. For the general case where Alice and Bob observe correlated $X, Y$ respectively, if we require an error probability less than $\epsilon$, the average key length can be at least
\[
I(X; Y)-2\log(I(X; Y)+1) - \log \lceil 1/\epsilon\rceil - 9.04.
\]

As a motivating example, consider the setting where Alice and Bob both observe $X$ with $p_X(0)=1/2$, $p_X(1)=p_X(2)=1/4$. They want to extract a key that contains i.i.d. fair coin flips. The best fixed-length key is $K_F=0$ if $X=0$, $K_F=1$ if $X\neq 0$ with length $1$. A variable-length key under the conditional uniformity requirement~\cite{li2021one} does not give any improvement since we cannot possibly have $L=2$, which requires the key to be uniform over $4$ values, considering that $X$ has only $3$ values. 
Consider a variable-length key given by $K_V=0$ if $X=0$, $K_V=10$ if $X=1$, $K_V=11$ if $X=2$, with expected length $3/2$. This can also be regarded as a (variable-length) sequence of fair coin flips, where we flip a coin to decide the first bit of the key, and then flip a second coin and append it to the key if the first coin is $1$.
We can argue that $K_V$ is better than the fixed-length key $K_F$. If it turns out that we only require a key of length $1$, then we can always take the length-$1$ prefix of $K_V$ (which is the same as $K_F$). Nevertheless, if we have a long i.i.d. sequence of such keys $K_{V,1},\ldots, K_{V,n}$, we can concatenate them to obtain a key of (almost fixed) length $3n/2$, whereas concatenating fixed-length keys $K_{F,1},\ldots, K_{F,n}$ will only yield a key of length $n$. We can see that the variable-length key $K_V$ is better in the sense that it retains all useful information of $K_F$, while giving a longer length under concatenation. In this sense, it is similar to Huffman coding, where variable-length codes can be concatenated to form a longer code.\footnote{An objection to using the variable-length key $K_V$ is that if its length $|K_V|$ is leaked to Eve, then Eve can tell the first bit of $K_V$. Nevertheless, if we concatenate the sequence of keys $K_{V,1},\ldots, K_{V,n}$ and take a fixed-length prefix, then the lengths of the individual keys will not be leaked.}

Note the similarity between the case where Alice and Bob observe a common $X$, and the Knuth-Yao random number generation scheme~\cite{knuth1976complexity}. In~\cite{knuth1976complexity}, it was shown that we can generate a random variable $X$ using an expected $H(X)+2$ number of coin flips (this is a variable-length scheme as the number of coin flips is not fixed). Here we show that we can generate a variable-length key (which can be regarded as a variable-length sequence of coin flips) of expected length $H(X)-2$ using $X$. The two results together show that we can convert between the random variable $X$ and $\approx H(X)$ number of coin flips, with a penalty of $2$ bits for each conversion. In this sense, these two results are ``dual'' to each other.

\iflongversion
Some proofs are moved to the appendix due to space constraints.
\else
Some proofs are moved to the preprint~\cite{zhou2024variable} due to space constraints.
\fi

\subsection*{Other Previous Works}

The study on randomness extractors~\cite{elias1972efficient,blum1986independent,zuckerman1990general,nisanRandomnessLinearSpace1996} concerns the setting where we extract an approximately uniformly distributed random number out of a random source with nonuniform distribution that may not be known exactly. It is similar to the setting in this paper where $X=Y$, though we always assume that the distribution of the source is known (but can be nonuniform). A closely-related line of research is universal hashing~\cite{carter1977universal}, which can be applied to enhance the privacy of secret keys by public discussion~\cite{bennett1988privacy, bennettGeneralizedPrivacyAmplification1995}. Dodis et al. \cite{dodisFuzzyExtractorsHow2008} constructed fuzzy extractors, based on randomness extractors, to generate uniformly distributed secret keys from observations of $X$ and its noisy version $X'$. Their results were further generalized in~\cite{fullerWhenAreFuzzy2020}.

\subsection*{Notations}

Logarithms are to the base $2$, and entropy is in bits. Write $\mathcal{K}^{*}:=\bigcup_{n=0}^{\infty}\mathcal{K}^{n}$ for
the set of sequences of entries in $\mathcal{K}$ with any length.
Given a sequence $K\in\mathcal{K}^{*}$, write $|K|$ for the length
of $K$, write $K_{i}$ for the $i$-th entry,
\[
K_{i}^{j}:=(K_{i},K_{i+1},\dots,K_{j})\in\mathcal{K}^{j-i+1},
\]
and $K^{j}:=K_{1}^{j}$. Write $K_1 \Vert K_2$ for the concatenation of two sequences. For an event $E$, its indicator function (which is $1$ if $E$ occurs, and $0$ otherwise) is written as $\mathbf{1}\{E\}$.
\medskip

\section{Randomly-Stopped Secret Key}

We say that a random $K\in\{0,1\}^{*}$ is a \emph{randomly-stopped
bit sequence} if
\[
\mathbb{P}\big( K_{n} = k_n \,\big|\,|K|\ge n,\,K^{n-1}=k^{n-1} \big) = 1/2
% K_{n}\,\big|\,\big\{|K|\ge n,\,K^{n-1}=k^{n-1}\big\}\,\sim\,\mathrm{Bern}(1/2)
\]
for all $n\ge1$ and $k^{n}\in\{0,1\}^{n}$ satisfying $\mathbb{P}(|K|\ge n,\,K^{n-1}=k^{n-1}) > 0$. The distribution
of such $K$ is called a randomly-stopped bit sequence distribution.
Intuitively, each bit $K_n$, if it exists (i.e., $|K|\ge n$), must be a fair coin flip independent of all past bits $K^{n-1}$.
Equivalently, $K$ has the same distribution as $B^M$, where $B_1, B_2,\ldots \stackrel{iid}{\sim} \mathrm{Bern}(1/2)$, and $M$ is a (possibly randomized) stopping time of $B_1, B_2,\ldots$. The readers are referred to~\cite{solan2012random} for an overview of randomized stopping time.

One example of a randomly-stopped bit sequence is given as follows.
\iflongversion
The proof is in Appendix~\ref{subsec:prefix}.
\else
The proof is in the preprint~\cite{zhou2024variable}.
\fi

\smallskip
\begin{prop}
\label{prop:prefixfree}Let $\mathcal{C} \subseteq \{0,1\}^*$ be a full prefix-free codebook (i.e., the corresponding coding tree is full, or equivalently, $\sum_{k \in \mathcal{C}}2^{-|k|} = 1$). Assume $K$ has the probability mass function $p_K(k)=2^{-|k|}$ for $k \in \mathcal{C}$. Then $K$ is a randomly-stopped
bit sequence.
\end{prop}
\smallskip

A useful property about randomly-stopped bit sequences is that their concatenations are also randomly-stopped
bit sequences. 
\iflongversion
The proof is given in Appendix \ref{subsec:concat}.
\else
The proof is in the preprint~\cite{zhou2024variable}.
\fi

\smallskip
\begin{prop}
\label{prop:concat}If $J,K$ are two independent randomly-stopped bit sequences, then $J \Vert K$ is a randomly-stopped bit sequence.
\end{prop}
\smallskip

We say that $K\in\{0,1\}^{*}$ is a \emph{conditionally randomly-stopped
bit sequence} given $W$ (where $K,W$ are jointly-distributed random
variables) if the conditional distribution $p_{K|W=w}$ is a randomly-stopped
bit sequence distribution for all $w$ in the support of $W$. Note that a conditionally randomly-stopped
bit sequence is also an (unconditionally) randomly-stopped
bit sequence, which follows from the definition.

We now describe the key agreement setting, which is similar to the setting in~\cite{li2021one}. Consider correlated random variables $X, Y$, over $\mathcal{X},\mathcal{Y}$, respectively. Let $G_A, G_B$ be random variables representing the local randomness for Alice and Bob, and $(X, Y), G_A, G_B$ are mutually independent. Alice observes $X$, and Bob observes $Y$. During the public discussion, Alice sends $W_1$ (dependent on $X, G_A$) to Bob, and then Bob sends $W_2$ (dependent on $Y, G_B, W_1$) to Alice, and then Alice sends $W_3$ to Bob, and so on, until they both agree to stop at time $N$ after sending $W_N$.
If $N=1$ always holds for a scheme, then we say that the scheme uses only one-way communication.
Finally, Alice outputs $K_{A}$ (dependent on $X, G_A, W^N$), and Bob outputs $K_{B}$ (dependent on $Y, G_B, W^N$). We say that
the key-agreement scheme that generates $(K_{A},K_{B})$ achieves
the error-length pair $(\epsilon,\ell)$ if there exists $K\in\{0,1\}^{*}$
(random and jointly distributed with all aforementioned random variables)
satisfying these three requirements:
\begin{itemize}
\item (Uniformity and secrecy). $K$ is a conditionally randomly-stopped bit sequence given $W^{N}$.
\smallskip{}
\item (Agreement probability). 
\[
\mathbb{P}\left(K_{A}=K_{B}=K\right)\ge1-\epsilon.
\]
%\smallskip{}
\item (Average length). 
\begin{align}
\mathbb{E}\left[|K|\cdot\mathbf{1}\{K_{A}=K_{B}=K\}\right]\ge\ell. \label{eq:errorlength}
\end{align}
\end{itemize}

Here $K$ is the ``ideal key'' that is exactly a conditionally randomly-stopped bit sequence given the public discussion $W^{N}$, i.e., it is a randomly-stopped sequence of fair coin flips regardless of the value of $W^{N}$. We require both actual keys $K_A, K_B$ to be close to the ideal $K$, which ensures that, according to an eavesdropper who observes $W^N$, the keys will always look like i.i.d. fair coin flips. Note that it is important that we consider $\mathbb{E}\left[|K|\cdot\mathbf{1}\{K_{A}=K_{B}=K\}\right]$ instead of $\mathbb{E}\left[|K|\right]$ (or $\mathbb{E}\left[|K_A|\right]$), since the latter can be made arbitrarily large by setting $K, K_A, K_B$ to be arbitrarily long with a very small probability of agreement. Therefore, we should count the length of the keys only when the keys agree.
We remark that, similar to~\cite{li2021one}, the key length $|K|$ does not need to be kept secret.

We first prove a converse bound, which shows that $\ell$ is upper-bounded by $I(X; Y)$ plus a constant.\footnote{Note that the bound in Proposition~\ref{prop:converse} does not depend on $\epsilon$. The average length $\ell$ cannot be too large regardless of the value of $\epsilon$.} 
\smallskip
\begin{prop}
\label{prop:converse}If the error-length pair $(\epsilon,\ell)$ is achievable, then
\[
\ell\le I(X;Y) + \log 3 + 1.
\]
\end{prop}
\begin{IEEEproof}
If $K$ is a randomly-stopped bit sequence, then
\begin{align*}
 & \mathbb{P}(K=k)\\
 & \le\mathbb{P}(|K|\ge|k|,\,K^{|k|}=k)\\
 & =\prod_{n=1}^{|k|}\mathbb{P}\big(|K|\ge n,\,K_{n}=k_{n}\,\big|\,|K|\ge n-1,\,K^{n-1}=k^{n-1}\big)\\
 & \le\prod_{n=1}^{|k|}\mathbb{P}\big(K_{n}=k_{n}\,\big|\,|K|\ge n,\,K^{n-1}=k^{n-1}\big) \\
 & = 2^{-|k|}.
\end{align*}
Therefore, if $K$ is a conditionally randomly-stopped bit sequence
given $W^{N}$, then $\mathbb{P}(K=k\,|\,W^{N}=w^{n})\le2^{-|k|}$ for any fixed $w^n$.
For $K\in\{0,1\}^{*}$, we write $K\cdot\mathbf{1}\{E\}\in\{0,1\}^{*}\cup\{0\}$,
where $K\cdot\mathbf{1}\{E\}=K$ if the event $E$ occurs, and $K\cdot\mathbf{1}\{E\}=0$
if the event $E$ does not occur, where we assume $0\notin\{0,1\}^{*}$
is a distinct symbol. We have
\begin{align}
 & H(K\cdot\mathbf{1}\{K=K_{A}=K_{B}\}\,|\,W^{N}=w^{n}) \nonumber \\
 & =-(1-\mathbb{P}(K=K_{A}=K_{B}|W^{N}=w^{n})) \nonumber \\
 & \;\;\;\;\;\;\;\cdot\log(1-\mathbb{P}(K=K_{A}=K_{B}|W^{N}=w^{n})) \nonumber \\
 & \;\;\;-\sum_{k\in\{0,1\}^{*}}\mathbb{P}(K=K_{A}=K_{B}=k\,|\,W^{N}=w^{n}) \nonumber \\
 & \;\;\;\;\;\;\;\cdot\log\mathbb{P}(K=K_{A}=K_{B}=k\,|\,W^{N}=w^{n}) \nonumber \\
 & \ge-\sum_{k\in\{0,1\}^{*}}\mathbb{P}(K=K_{A}=K_{B}=k\,|\,W^{N}=w^{n}) \nonumber \\
 & \;\;\;\;\;\;\;\cdot\log\mathbb{P}(K=k\,|\,W^{N}=w^{n}) \nonumber \\
 & \ge\sum_{k\in\{0,1\}^{*}}\mathbb{P}(K=K_{A}=K_{B}=k\,|\,W^{N}=w^{n})\cdot|k| \nonumber \\
 & =\mathbb{E}\big[|K|\cdot\mathbf{1}\{K=K_{A}=K_{B}\}\,\big|\,W^{N}=w^{n}\big]. \label{eq:ent_K}
\end{align}
Invoking Lemma 2 in \cite{li2021one}, which gives $\mathbb{P}(K_{A}=K_{B})H(K_{A}|W^{N},K_{A}=K_{B})\le I(X;Y)+1$,
we have
\begin{align*}
 & I(X;Y)+1\\
 & \ge\mathbb{P}(K_{A}=K_{B})H(K_{A}|W^{N},K_{A}=K_{B})\\
 & =H(K_{A}\cdot\mathbf{1}\{K_{A}=K_{B}\}\,|\,W^{N},\mathbf{1}\{K_{A}=K_{B}\})\\
 & \ge H(K_{A}\cdot\mathbf{1}\{K_{A}=K_{B}\},\,\mathbf{1}\{K=K_{A}=K_{B}\}\,|\,W^{N})\\
 & \;\;\;\;-H(\mathbf{1}\{K_{A}=K_{B}\},\,\mathbf{1}\{K=K_{A}=K_{B}\})\\
 & \ge H(K\cdot\mathbf{1}\{K=K_{A}=K_{B}\}\,|\,W^{N})-\log3\\
 & \ge \mathbb{E}\big[|K|\cdot\mathbf{1}\{K=K_{A}=K_{B}\}\big|W^{N}\big]-\log3,
\end{align*}
where the last inequality is by \eqref{eq:ent_K}.
\end{IEEEproof}

A useful property of randomly-stopped secret keys is that they can be concatenated to form longer keys. Therefore, to convert from several randomly-stopped secret keys to a conventional fixed-length key, we can concatenate the randomly-stopped secret keys, giving a key with length concentrated around its mean by the law of large numbers, and then take a fixed-length prefix of that key. 
\smallskip
\begin{prop}
Assume that $(K_{A,i},K_{B,i})$ is a randomly-stopped secret key with public discussion $W_i^{N_i}$ achieving the error-length pair $(\epsilon_i,\ell_i)$ for $i=1,2$, and $(K_{A,1},K_{B,1},W_1^{N_1})$ is independent of $(K_{A,2},K_{B,2},W_1^{N_2})$. Then the concatenation $(K_{A,1}\Vert K_{A,2},K_{B,1} \Vert K_{B,2})$ is a randomly-stopped secret key with public discussion $(W_1^{N_1},W_2^{N_2})$ achieving the error-length pair $(\epsilon_1 + \epsilon_2,\, (1-\epsilon_2)\ell_1 + (1-\epsilon_1)\ell_2)$.
\end{prop}
\begin{IEEEproof}
Let $K_i$ be the conditionally randomly-stopped bit sequence in \eqref{eq:errorlength}. By Proposition~\ref{prop:concat}, $K_1\Vert K_2$ is a conditionally randomly-stopped bit sequence given $(W_1^{N_1},W_2^{N_2})$. We have 
\begin{align*}
& 1-\mathbb{P}(K_{A,1}\Vert K_{A,2} =K_{B,1}\Vert K_{B,2}=K_{1}\Vert K_{2})\\
& \le 1-\mathbb{P}(K_{A,1} =K_{B,1}=K_{1}) + 1-\mathbb{P}(K_{A,2} =K_{B,2}=K_{2})\\
& \le \epsilon_1 + \epsilon_2,
\end{align*}
and
\begin{align*}
& \mathbb{E}\left[(|K_1| +| K_2|)\cdot\mathbf{1}\{K_{A,1}\Vert K_{A,2} =K_{B,1}\Vert K_{B,2}=K_{1}\Vert K_{2}\}\right] \\
& \le \mathbb{P}(K_{A,2} =K_{B,2}=K_{2}) \mathbb{E}\left[|K_1|\cdot\mathbf{1}\{K_{A,1} =K_{B,1}=K_{1}\}\right]\\
&\;\; + \mathbb{P}(K_{A,1} =K_{B,1}=K_{1}) \mathbb{E}\left[|K_2|\cdot\mathbf{1}\{K_{A,2} =K_{B,2}=K_{2}\}\right] \\
& \le (1-\epsilon_2)\ell_1 + (1-\epsilon_1)\ell_2.
\end{align*}
\\[-2.5\baselineskip]
\end{IEEEproof}

In the remaining section, we will design schemes for three cases: 1) $X=Y$ always holds, 2) $X=Y$ holds with high probability, and 3) general correlated $X, Y$.

\section{Common Randomness Model}

We start with the simple case $X=Y$. This setting is useful when Alice and Bob share a common random source with a nonuniform distribution, and want to extract uniformly distributed bits. Before we present the scheme, we introduce a useful lemma. 
A \emph{dyadic distribution} $p$ is a distribution where each nonzero $p(x)$
is a power of two.
\smallskip
\begin{lem}\label{lem:dyadic}
For any discrete random variable $X$, there exists $p_{W|X}$,
where $W$ is a random variable with marginal distribution $W\sim\mathrm{Geom}(1/2)$,  and $p_{X|W=w}$ is a dyadic distribution
for all $w \ge 1$.
\end{lem}
\begin{IEEEproof}
We will first prove the following claim: there exists an event $E$ such that $\mathbb{P}(E)=1/2$ and $p_{X|E}$ is a dyadic distribution.
Without loss of generality, assume $p_X(1) \ge p_X(2) \ge \cdots$. The case $p_X(1)=1$ is obvious, so we further assume $p_X(1)<1$. For each $x$, let $\alpha_x$ be the smallest integer satisfying $2^{-\alpha_x}\le p_X(x)$. 
Since $2^{-\alpha_x}> p_X(x)/2$, we have $\sum_{x=1}^{\infty}2^{-\alpha_x}> 1/2$.
Let $x_0$ be the largest integer such that $\sum_{x=1}^{x_0}2^{-\alpha_x}\le 1/2$. We will prove that $\sum_{x=1}^{x_0}2^{-\alpha_x}= 1/2$. Assume the contrary that $\sum_{x=1}^{x_0}2^{-\alpha_x}< 1/2$. Note that since $\alpha_x$ is nondecreasing in $x$, $2^{-\alpha_x}$ is a multiple of $2^{-\alpha_{x_0}}$ for $x \le x_0$. Since $1/2$ is also a multiple of $2^{-\alpha_{x_0}}$, $1/2 - \sum_{x=1}^{x_0}2^{-\alpha_x} > 0$ is a multiple of $2^{-\alpha_{x_0}}$, and hence is a multiple of $2^{-\alpha_{x_0+1}}$. This implies that $2^{-\alpha_{x_0+1}} \le 1/2 - \sum_{x=1}^{x_0}2^{-\alpha_x}$, $\sum_{x=1}^{x_0+1}2^{-\alpha_x}\le 1/2$, contradicting with the maximality of $x_0$. Therefore, we have $\sum_{x=1}^{x_0}2^{-\alpha_x}= 1/2$. Let $\mathbb{P}(E|X=x)=2^{-\alpha_x}/p_X(x)$. We have $\mathbb{P}(E)=1/2$ and $p_{X|E}(x)=2^{1-\alpha_x}$. The claim follows.

We now complete the proof of the lemma. Let $E_1, E_2,\ldots$ be constructed recursively. Applying the claim on $p_{X|\bigcap_{j=1}^{i-1} E_j^c}$, we can construct $E_i \subseteq \bigcap_{j=1}^{i-1} E_j^c$ such that $\mathbb{P}(E_i | \bigcap_{j=1}^{i-1} E_j^c)=1/2$ and $p_{X|E_i}=p_{X|E_i\cap \bigcap_{j=1}^{i-1} E_j^c}$ is a dyadic distribution. Note that $\mathbb{P}(E_i)=2^{-i}$, and hence $E_1, E_2,\ldots$ partition the probability space (except a null set). Let $W$ be an integer such that the event $E_W$ occurs. The result follows.
\end{IEEEproof}

We use the lemma to construct a scheme for the case $X=Y$ with zero error, with an expected length of at least $H(X)-2$. 
\smallskip
\begin{thm}
\label{thm:common}Consider $X=Y$. There is a scheme that uses only one-way communication achieving the error-length pair $(0,\, H(X)-2)$.
\end{thm}
\begin{IEEEproof}
By Lemma~\ref{lem:dyadic}, let $W$ be a random variable with marginal distribution $W\sim\mathrm{Geom}(1/2)$ such that $p_{X|W=w}$ is a dyadic distribution. For each $w$, by Kraft's inequality, there exists a prefix-free code $f_{w}: \mathcal{X} \to \{0,1\}^*$ such that $|f_w(x)| = -\log(p_{X|W=w}(x))$. Alice generates $W$ conditioned on $X$, and sends $W$ to Bob. Then, both Alice and Bob output $K=f_W(X)$, which is a conditionally randomly-stopped bit sequence given $W$ by Proposition~\ref{prop:prefixfree}. We have 
\begin{align*}
& \mathbb{E}[|K|] \,=\, \mathbb{E}\big[\big|f_W(X) \, \big|\big] \,=\, \mathbb{E}\big[-\log(p_{X|W}(X|W)) \, \big]\\
 &= H(X|W) \,\ge\, H(X)-H(W) \, =\, H(X)-2.
\end{align*}
\\[-2.5\baselineskip]
\end{IEEEproof}

The key agreement algorithm, which follows the construction in the
proofs of Lemma~\ref{lem:dyadic} and Theorem \ref{thm:common}, is given in Algorithm \ref{alg:keyagree}. The correctness
of the algorithm follows from the proofs of the lemma and theorem. Note that the
expected total number of iterations of the outer for loop is $2$
since $W \sim \mathrm{Geom}(1/2)$. Assuming floating point
operation is constant time, the time complexity of the algorithm is
$O(|\mathcal{X}|\log|\mathcal{X}|)$ due to the sorting step.

\begin{algorithm}%[H]
\iflongversion
\textbf{$\;\;\;\;$Input:} $\mathrm{id}\in\{\mathrm{Alice},\mathrm{Bob}\}$,
probability mass function $p$, value $x\in\{1,\ldots,|\mathcal{X}|\}$,
discussion $w\in\mathbb{N}$ (for Bob)
\else
\textbf{$\;\;\;\;$Input:} $\mathrm{id}\in\{\mathrm{Alice},\mathrm{Bob}\}$,
probability mass function $p$, 

\textbf{$\;\;\;\;$$\;\;\;\;$}value $x\in\{1,\ldots,|\mathcal{X}|\}$,
discussion $w\in\mathbb{N}$ (for Bob)
\fi

\textbf{$\;\;\;\;$Output:} key $k\in\{0,1\}^{*}$, discussion $w\in\mathbb{N}$
(for Alice)

\smallskip{}

\begin{algorithmic}

\State{$\tilde{p}\leftarrow p$}

\If{$\mathrm{id}=\mathrm{Alice}$}

\State{generate $g\sim\mathrm{Unif}[0,p(x)]$}

\EndIf

\For{$w'=1,2,\ldots$}

\State{$q\leftarrow2^{-w'}$, $k\leftarrow0$}

\State{let $y_{1},\ldots,y_{|\mathcal{X}|}$ such that $\tilde{p}(y_{1})\ge\cdots\ge\tilde{p}(y_{|\mathcal{X}|})$}

\For{$i=1,\ldots,|\mathcal{X}|$}

\State{$\alpha\leftarrow\max\{\lceil-\log\tilde{p}(y_{i})\rceil,\,w'\}$}

\If{$2^{-\alpha}>q$}

\State{$\textbf{break for}$}

\EndIf

\State{$q\leftarrow q-2^{-\alpha}$, $\; \tilde{p}(y_{i})\leftarrow\tilde{p}(y_{i})-2^{-\alpha}$}

\If{$y_{i}=x$}

\If{$\mathrm{id}=\mathrm{Alice}$ and $g\ge\tilde{p}(y_{i})$}

\iflongversion
\State{$\textbf{return}$ first $\alpha-w'$ binary digits of $k$ (after the decimal point), and $w'$}
\else
\State{$\textbf{return}$ first $\alpha-w'$ binary digits of $k$
}

\State{$\qquad\quad$(after the decimal point), and $w'$}
\fi

\ElsIf{$\mathrm{id}=\mathrm{Bob}$ and $w=w'$}

\State{$\textbf{return}$ first $\alpha-w'$ binary digits of $k$}

\EndIf

\EndIf

\State{$k\leftarrow k+2^{-(\alpha-w')}$}

\EndFor

\EndFor

\end{algorithmic}

\caption{\label{alg:keyagree}$\textsc{KeyAgree}(\mathrm{id},p,x,w)$}
\end{algorithm}

\section{Almost Common Randomness Model}
We then consider the case where $X$ and $Y$ agree with probability $p:=\mathbb{P}(X=Y)$, and $p H(X|X=Y)$ is large. 

\smallskip
\begin{thm}
\label{thm:entropy}Let $p:=\mathbb{P}(X=Y)$. For any integer $m\ge 1$, there is a scheme using $2$ rounds of interactive communication achieving the error-length pair $(\epsilon, \ell)$, where
\begin{align*}
\epsilon & = (1-p)/m, \\ % \frac{1-p}{m}, \\
\ell &= p(H(X|X=Y)-\log {m} -2).
\end{align*}
\end{thm}
\smallskip

Note the similarity between this result and the entropy model in~\cite{li2021one}, where it was shown that an expected length
\[
\approx \kappa - \log \kappa - 2 \log (1/\epsilon)
\]
is achievable under the conditional uniformity requirement in~\cite{li2021one}, where $\kappa := pH(X|X=Y)$. In comparison, Theorem~\ref{thm:entropy} shows that an expected length at least
\[
\ell = \kappa - \log \lceil1/\epsilon \rceil - 2
\]
is achievable under the new conditionally randomly-stopped bit sequence requirement, by taking $m=\lceil1/\epsilon \rceil$. We can see that the new setting allows a longer key without the $\log \kappa$ term. In addition, the almost common randomness model helps to develop the general key agreement scheme in the next section. We now prove the theorem.
\begin{IEEEproof}
Fix $m \ge 1$. Alice generates $F:\mathcal{X}\cup \mathcal{Y}\rightarrow \{1, \dots, m\}$, a random function uniformly sampled from the space of functions from $\mathcal{X}\cup \mathcal{Y}$ to $\{1, \dots, m\}$, which will serve as a hash function. 
Alice sends $W_1=(F, F(X))$ to Bob (we will see later that $F$ does not need to be sent). If $F(X)=F(Y)$ holds, Bob will proceed with the scheme in Theorem~\ref{thm:common} applied on the conditional probability distribution $p_{X|X=Y, F(X), F}(\,\cdot \, | W_1, F)$ to output the one-way communication $W_2$ from Bob to Alice, in order to allow them generate the keys $K_A$ and $K_B$.
%This is valid as we have $\mathbb{P}(X\neq Y|W_1=W_2)=\frac{1-\mathbb{P}(X=Y)}{1+(m-1)\mathbb{P}(X=Y)}= \frac{1-p}{1+(m-1)p}$. 
Otherwise, Bob sends $W_2 = \mathrm{e}$ (an error symbol) to Alice, and Alice and Bob will output the empty key $K_A=K_B= \emptyset$. 
% We can see that both $K_A$ and $K_B$ are conditionally randomly-stopped bit sequences given $(W_1,W_2,W_3)$.

First we bound the expected key length when $X=Y$, which guarantees $F(X)=F(Y)$. By Theorem~\ref{thm:common},
\begin{align*}
&\mathbb{E}\big[|K_A| \, \big| \,X=Y,\, F=f \big]\\
& = \mathbb{E}\big[\mathbb{E}\big[|K_A| \, \big| \,X=Y,\, F=f ,\, F(X) \big] \, \big| \, X=Y,\, F=f \big]\\
&\ge H(X|X=Y,F(X),F=f)-2 \\
&= H(X|X=Y,f(X))-2 \\
&= H(X|X=Y)-H(f(X)|X=Y)-2\\&
\ge H(X|X=Y)-\log {m} -2. 
\end{align*}
% \begin{align*}
% &\mathbb{E}\big[|K_A| \, \big| \,X=Y \big]\\
% & = \mathbb{E}\big[\mathbb{E}\big[|K_A| \, \big| \,X=Y, F(X),F \big] \big]\\
% &\ge H(X|X=Y,F(X),F)-2 \\
% &= H(X|X=Y,F)-H(F(X)|X=Y,F)-2\\&
% \ge H(X|X=Y)-\log {m} -2. 
% \end{align*}
Let $K=K_A=K_B$ if $X=Y$, and $K=\emptyset$ if $X\neq Y$. By construction, $K$ is a conditionally randomly-stopped bit sequence given $(W_1, W_2)$.
We have
\begin{align}
&\mathbb{E}\big[|K| \cdot \mathbf{1}\{K_A=K_B=K\} \,|\, F=f \big] \nonumber\\
& = \mathbb{E}\big[|K_A| \cdot  \mathbf{1}\{X=Y\}  \,|\, F=f \big] \nonumber\\
&\ge p(H(X|X=Y)-\log {m} -2). \label{eq:bound_len}
\end{align}
We then bound the error probability. Since we have $K_A=K_B=K$ if $X=Y$ or $F(X) \neq F(Y)$, we have
\begin{align}
& 1 - \mathbb{P}(K_A = K_B = K) \,\le\, \mathbb{P}(F(X)=F(Y) , \, X \neq Y)  \nonumber\\
&= (1-p) \mathbb{P}(F(X)=F(Y) \, | \, X \neq Y)   \,=\, (1-p)/m \label{eq:bound_pe}
\end{align}
since $F$ is a random function. The result follows.

We now argue that $F$ can be fixed, and does not need to be sent via public discussion. By \eqref{eq:bound_pe}, there exists a fixed $f$ achieving $\mathbb{P}(K_A \neq K_B \, | \, F=f) \le (1-p)/m$. Since \eqref{eq:bound_len} holds conditioned on any choice of $F$, the bound on the length is valid conditioned on $f$ as well. Therefore, we can fix $F=f$, eliminating the need to send it.
\end{IEEEproof}

In practice, we will fix $F=f$ to be a 
hash function. After Alice sends $w_1=f(x)$, Bob will run Algorithm~\ref{alg:keyagree} on the conditional distribution $p_{X|X=Y, f(X)=w_1}$ if $w_1=f(y)$ (note that the roles of Alice and Bob are swapped here compared to Algorithm~\ref{alg:keyagree}), or output the empty key if $w_1 \neq f(y)$. Assuming $p_{X|X=Y}$ is provided to Alice and Bob, and hash values can be computed in constant time, it takes $O(|\mathcal{X}|)$ to compute $p_{X|X=Y, f(X)=w_1}$, so the overall complexity for the key agreement scheme is still $O(|\mathcal{X}|\log|\mathcal{X}|)$.

\section{Correlated Randomness Model}
We finally consider the case for $X$ and $Y$ sampled from a general random source. We will invoke \cite[Lemma 2]{li2021one}, which states that there exists a scheme which allows Alice and Bob to generate $M_A$ (function of $X, G_A, W^N$) and $M_B$ (function of $Y, G_B, W^N$) respectively, after the public discussion $W_1, W_2, \dots, W_N$, satisfying
\begin{align}
&\mathbb{P}(M_A = M_B) H(M_A | W^N, M_A=M_B) \nonumber \\
&\ge I(X;Y)-2 \log (I(X;Y)+1)-7.04.\label{eq:kappa}
\end{align}

Our scheme works as follows. First, Alice and Bob use the scheme in \cite[Lemma 2]{li2021one} to produce $M_A$ and $M_B$ respectively, using public discussion $W_1, W_2, \dots, W_N$. Then they proceed with the scheme in Theorem~\ref{thm:entropy} applied on the conditional distribution $p_{M_A,M_B | W^N}$ to generate the key $K_A$ and $K_B$, using public discussion $W_{N+1}, W_{N+2}$.

\smallskip
\begin{thm}
\label{thm:general} 
% For any integer $m\ge 1$, there is a scheme using $2$ rounds of interactive communication achieving the error-length pair $(\epsilon, \ell)$, where
Fix any $m \ge 1$. The above protocol achieves the error-length pair $(\epsilon, \ell)$, where
\begin{align*}
\epsilon & = 1/m, \\
\ell &= I(X; Y)-2\log(I(X; Y)+1)-\log m - 9.04.
\end{align*}
\end{thm}

\begin{IEEEproof}
Consider the case $W^N=w^n$. Applying Theorem~\ref{thm:entropy} on $p_{X,Y} \leftarrow p_{M_A,M_B | W^N=w^n}$, Alice and Bob can then generate keys $K_A$ and $K_B$ respectively using public discussion $W_{N+1}, W_{N+2}$, and there exists $K$ which is a conditionally randomly-stopped bit sequence given $(W_{N+1}, W_{N+2})$ and $W^N=w^n$, satisfying $\mathbb{P}(K_A=K_B=K\, | \, W^N=w^n) \ge 1-1/m$ and 
\begin{align*}
& \mathbb{E}[|K|\mathbf{1}\{ K_A=K_B=K \}\, | \, W^N=w^n] \\
& \ge \mathbb{P}(M_A = M_B | W^N=w^n) H(M_A | W^N=w^n, M_A=M_B) \\
& \;\;\;\;- \log m -2.
\end{align*}
Averaging over $w^n$, we have $\mathbb{P}(K_A=K_B=K) \ge 1-1/m$, and by~\eqref{eq:kappa},
\begin{align*}
& \mathbb{E}[|K|\mathbf{1}\{ K_A=K_B=K \}] \\
& \ge \sum_{w^n} \mathbb{P}(W^N=w^n) \mathbb{P}(M_A = M_B | W^N=w^n) \\
& \;\;\;\;\; \cdot H(M_A | W^N=w^n, M_A=M_B) - \log m -2 \\
& = \mathbb{P}(M_A = M_B) \sum_{w^n} \mathbb{P}(W^N=w^n | M_A = M_B) \\
& \;\;\;\;\; \cdot H(M_A | W^N=w^n, M_A=M_B) - \log m -2 \\
& = \mathbb{P}(M_A = M_B) H(M_A | W^N, M_A=M_B) - \log m -2 \\
& \ge I(X;Y)-2 \log (I(X;Y)+1) - \log m -9.04.
\end{align*}
\\[-2.5\baselineskip]
\end{IEEEproof}
This theorem, together with Proposition~\ref{prop:converse}, shows that the optimal average length is within a logarithmic gap from $I(X; Y)$.
Proposition~\ref{prop:converse} and Theorem~\ref{thm:general} imply that, if we consider a general source $(\mathbf{X}_n,\mathbf{Y}_n)_{n \ge 1}$ (i.e., each $(\mathbf{X}_n,\mathbf{Y}_n)$ is a pair of random variables that are not necessarily i.i.d. sequences) with mutual information rate $\lim_{n\to \infty}I(\mathbf{X}_n;\mathbf{Y}_n)/n = R$, then for any fixed agreement probability $\epsilon \in (0,1)$, letting $\ell_n$ be an achievable average length for the pair $(\mathbf{X}_n,\mathbf{Y}_n)$, we know that the largest possible key rate $\lim\inf_{n\to \infty}\ell_n/n$ is $R$.

\section{Conclusion}

We presented a new definition of variable-length secret keys based on randomized stopping time, and showed that it allows a longer secret key compared to the definition in~\cite{li2021one}. While the key length in the general case (Theorem~\ref{thm:general}) contains a logarithmic gap, it is left for future studies to investigate whether the logarithmic gap can be eliminated.

% \section{Acknowledgements}

% This work was partially supported by two grants from the Research Grants Council of the Hong Kong Special Administrative Region, China [Project No.s: CUHK 24205621 (ECS), CUHK 14209823 (GRF)].

\iflongversion
% \clearpage

\appendices
\section{Proof of Proposition~\ref{prop:prefixfree}\label{subsec:prefix}}

Note that $\mathbb{P}(|K|\ge n,\,K^{n}=k^{n}) > 0$ if and only if $k^{n}$ is a (possibly improper) prefix of some $c\in \mathcal{C}$. If $k^{n}$ is a prefix of some $c\in \mathcal{C}$, then 
\begin{align*}
& \mathbb{P}(|K|\ge n,\,K^{n}=k^{n}) \\
& = \sum_{c \in \mathcal{C}:\, c^n=k^n} 2^{-|c|} \\
& = 2^{-n} \sum_{c \in \mathcal{C}:\, c^n=k^n} 2^{-(|c| - n)} \\
& = 2^{-n}
\end{align*}
since $\{c_{n+1}^{|c|}:\, c \in \mathcal{C}:\, c^n=k^n\}$ is also a full prefix-free codebook, corresponding to the subtree with the root $k^n$ in the coding tree.

Also note that $\mathbb{P}(|K|\ge n,\,K^{n-1}=k^{n-1}) > 0$ if and only if $k^{n-1}$ is a proper prefix of some $c\in \mathcal{C}$. If $k^{n-1}$ is a proper prefix of some $c\in \mathcal{C}$, since the code is full, each of $k^{n-1}\Vert 0$ and $k^{n-1}\Vert 1$ is a prefix of some codewords. So we have $\mathbb{P}(|K|\ge n, K^n=k^{n-1}\Vert k_n)=2^{-n}$ for $k_n=0,1$. Therefore, $\mathbb{P}\big(K_n=k_n\big||K|\ge n,K^{n-1}=k^{n-1}\big)=1/2$ for $k_n=0,1$. Hence such $K$ is a randomly-stopped bit sequence.

\section{Proof of Proposition~\ref{prop:concat}\label{subsec:concat}}
Consider $B_1, B_2,\ldots \stackrel{iid}{\sim} \mathrm{Bern}(1/2)$ independent of $G_1, G_2,\ldots \stackrel{iid}{\sim} \mathrm{Unif}[0,1]$.
Let $\tilde{B}_i = (B_i, G_i)$.
We first show that $K$ is a randomly-stopped bit sequence if and only if there exists a stopping time $M$ of the process $\tilde{B}_1,\tilde{B}_2,\ldots$ such that $K$ has the same distribution as $B^M$.
For the ``if'' direction, since whether $M\ge n$ depends only on $(B^{n-1},G^{n-1})$, we have $\mathbb{P}( B_{n} = b_n \,|\,M\ge n,\,B^{n-1}=b^{n-1} ) = 1/2$. For the ``only if'' direction, we can take
\[
M = \min\big\{n:\, G_n \ge \rho(B^{n})  \big\},
\]
where 
\[
\rho(b^n):=\mathbb{P}\big( |K|\ge n+1 \,\big|\,|K|\ge n,\,K^{n}=b^{n} \big).
\]
It is straightforward to check $\mathbb{P}( B_{n} = b_n \,|\,M\ge n,\,B^{n-1}=b^{n-1} ) = 1/2$ and $\mathbb{P}( M\ge n \,|\,B^{n-1}=b^{n-1} ) = \rho(b^{n-1}) = \mathbb{P}( |K|\ge n \,|\,K^{n-1}=b^{n-1} )$, and hence $B^M$ has the same distribution as $K$.

To show that the concatenation of two independent randomly-stopped bit sequences $J,K$ is still a randomly-stopped bit sequence, 
assume $J$ has the same distribution as $B^N$, where $N=N((\tilde{B}_n)_{n \ge 1})$ (a function of $(\tilde{B}_n)_{n \ge 1}$) is the aforementioned stopping time. Also assume $K$ has the same distribution as $B^M$, where $M=M((\tilde{B}_n)_{n \ge 1})$ is the aforementioned stopping time. Consider the stopping time 
\[
L((\tilde{B}_n)_{n \ge 1}) := N((\tilde{B}_n)_{n \ge 1}) + M((\tilde{B}_{n+N((\tilde{B}_n)_{n \ge 1})})_{n \ge 1}),
\]
i.e., after we reach time $N$, we treat the remainder $(\tilde{B}_{n+N((B_n)_{n \ge 1})})_{n \ge 1}$ of the process as a new process and compute the stopping time $M$ on it, and stop. Note that $B_{N+1}^{L}$ has the same distribution as $B^M$ and $K$, and is independent of $B^N$ by the strong Markov property. Hence, $J\Vert K$ has the same distribution as $B^N \Vert B_{N+1}^{L} = B^{L}$, which is a randomly-stopped bit sequence.

\else

\fi

\iflongversion
\else
\IEEEtriggeratref{11}
\fi

\bibliographystyle{IEEEtran}
\bibliography{ref,paper}

\begin{thebibliography}{10}
\providecommand{\url}[1]{#1}
\csname url@samestyle\endcsname
\providecommand{\newblock}{\relax}
\providecommand{\bibinfo}[2]{#2}
\providecommand{\BIBentrySTDinterwordspacing}{\spaceskip=0pt\relax}
\providecommand{\BIBentryALTinterwordstretchfactor}{4}
\providecommand{\BIBentryALTinterwordspacing}{\spaceskip=\fontdimen2\font plus
\BIBentryALTinterwordstretchfactor\fontdimen3\font minus
  \fontdimen4\font\relax}
\providecommand{\BIBforeignlanguage}[2]{{%
\expandafter\ifx\csname l@#1\endcsname\relax
\typeout{** WARNING: IEEEtran.bst: No hyphenation pattern has been}%
\typeout{** loaded for the language `#1'. Using the pattern for}%
\typeout{** the default language instead.}%
\else
\language=\csname l@#1\endcsname
\fi
#2}}
\providecommand{\BIBdecl}{\relax}
\BIBdecl

\bibitem{gacsCommonInformationFar1973}
P.~G{\'a}cs and J.~K{\"o}rner, ``Common information is far less than mutual
  information,'' \emph{Problems of Control and Information Theory}, vol.~2,
  no.~2, pp. 149--162, 1973.

\bibitem{maurerSecretKeyAgreement1993}
U.~Maurer, ``Secret key agreement by public discussion from common
  information,'' \emph{IEEE Transactions on Information Theory}, vol.~39,
  no.~3, pp. 733--742, 1993.

\bibitem{ahlswedeCommonRandomnessInformation1993}
R.~Ahlswede and I.~Csiszar, ``Common randomness in information theory and
  cryptography. {{I}}. {{Secret}} sharing,'' \emph{IEEE Transactions on
  Information Theory}, vol.~39, no.~4, pp. 1121--1132, Jul. 1993.

\bibitem{maurerUnconditionallySecureKey1999}
U.~Maurer and S.~Wolf, ``Unconditionally secure key agreement and the intrinsic
  conditional information,'' \emph{IEEE Transactions on Information Theory},
  vol.~45, no.~2, pp. 499--514, Mar. 1999.

\bibitem{maurerInformationTheoreticKeyAgreement2000}
------, ``Information-theoretic key agreement: {{From}} weak to strong secrecy
  for free,'' in \emph{Advances in {{Cryptology}} \textemdash{} {{EUROCRYPT}}
  2000}, ser. Lecture {{Notes}} in {{Computer Science}}, B.~Preneel, Ed.\hskip
  1em plus 0.5em minus 0.4em\relax {Berlin, Heidelberg}: {Springer}, 2000, pp.
  351--368.

\bibitem{hayashiSecretKeyAgreement2016}
M.~Hayashi, H.~Tyagi, and S.~Watanabe, ``Secret key agreement: {{General}}
  capacity and second-order asymptotics,'' \emph{IEEE Transactions on
  Information Theory}, vol.~62, no.~7, pp. 3796--3810, Jul. 2016.

\bibitem{tyagi2015converses}
H.~Tyagi and S.~Watanabe, ``Converses for secret key agreement and secure
  computing,'' \emph{IEEE Transactions on Information Theory}, vol.~61, no.~9,
  pp. 4809--4827, Sept 2015.

\bibitem{tyagi2017universal}
------, ``Universal multiparty data exchange and secret key agreement,''
  \emph{IEEE Transactions on Information Theory}, vol.~63, no.~7, pp.
  4057--4074, 2017.

\bibitem{li2021one}
C.~T. Li and V.~Anantharam, ``One-shot variable-length secret key agreement
  approaching mutual information,'' \emph{IEEE Transactions on Information
  Theory}, vol.~67, no.~8, pp. 5509--5525, 2021.

\bibitem{knuth1976complexity}
D.~E. Knuth and A.~C. Yao, ``The complexity of nonuniform random number
  generation,'' \emph{Algorithms and Complexity: New Directions and Recent
  Results}, pp. 357--428, 1976.

\bibitem{elias1972efficient}
P.~Elias, ``The efficient construction of an unbiased random sequence,''
  \emph{The Annals of Mathematical Statistics}, vol.~43, no.~3, pp. 865--870,
  1972.

\bibitem{blum1986independent}
M.~Blum, ``Independent unbiased coin flips from a correlated biased source—a
  finite state {M}arkov chain,'' \emph{Combinatorica}, vol.~6, pp. 97--108,
  1986.

\bibitem{zuckerman1990general}
D.~Zuckerman, ``General weak random sources,'' in \emph{Proceedings [1990] 31st
  Annual Symposium on Foundations of Computer Science}.\hskip 1em plus 0.5em
  minus 0.4em\relax IEEE, 1990, pp. 534--543.

\bibitem{nisanRandomnessLinearSpace1996}
N.~Nisan and D.~Zuckerman, ``Randomness is linear in space,'' \emph{Journal of
  Computer and System Sciences}, vol.~52, no.~1, pp. 43--52, Feb. 1996.

\bibitem{carter1977universal}
J.~L. Carter and M.~N. Wegman, ``Universal classes of hash functions,'' in
  \emph{Proceedings of the ninth annual ACM symposium on Theory of computing},
  1977, pp. 106--112.

\bibitem{bennett1988privacy}
C.~H. Bennett, G.~Brassard, and J.-M. Robert, ``Privacy amplification by public
  discussion,'' \emph{SIAM journal on Computing}, vol.~17, no.~2, pp. 210--229,
  1988.

\bibitem{bennettGeneralizedPrivacyAmplification1995}
C.~Bennett, G.~Brassard, C.~Crepeau, and U.~Maurer, ``Generalized privacy
  amplification,'' \emph{IEEE Transactions on Information Theory}, vol.~41,
  no.~6, pp. 1915--1923, Nov. 1995.

\bibitem{dodisFuzzyExtractorsHow2008}
Y.~Dodis, R.~Ostrovsky, L.~Reyzin, and A.~Smith, ``Fuzzy {{Extractors}}:
  {{How}} to {{Generate Strong Keys}} from {{Biometrics}} and {{Other Noisy
  Data}},'' \emph{SIAM Journal on Computing}, vol.~38, no.~1, pp. 97--139, Jan.
  2008.

\bibitem{fullerWhenAreFuzzy2020}
B.~Fuller, L.~Reyzin, and A.~Smith, ``When are fuzzy extractors possible?''
  \emph{IEEE Transactions on Information Theory}, vol.~66, no.~8, pp.
  5282--5298, Aug. 2020.

\bibitem{solan2012random}
E.~Solan, B.~Tsirelson, and N.~Vieille, ``Random stopping times in stopping
  problems and stopping games,'' \emph{arXiv preprint arXiv:1211.5802}, 2012.

\end{thebibliography}

\end{document}